\documentclass[a4paper,12pt]{article}
\usepackage[a4paper,text={16.8cm,22.4cm}]{geometry}
\usepackage{amsmath,amssymb,latexsym,bm,braket}
\usepackage{epsfig,psfrag,color}
\usepackage{graphicx}

\usepackage{cite}

\allowdisplaybreaks 
\usepackage{color}

\newcommand{\nn}{\nonumber}

\newcommand{\be}{\begin{equation}}
\newcommand{\ee}{\end{equation}}
\newcommand{\bea}{\begin{eqnarray}}
\newcommand{\eea}{\end{eqnarray}}

\newcommand{\msbar}{\overline{\text{MS}}}

\begin{document}

\begin{titlepage}

  \begin{flushright}
  IPPP/14/81 \\
  DCPT/14/162\\
  PSI-PR-14-11
  \end{flushright}
  
  \vspace{5ex}
  
  \begin{center}
    \boldmath
    \textbf{\Large NNLO hard functions in massless QCD} \vspace{7ex}
    \unboldmath    
    
    \textsc{Alessandro Broggio$^a$, Andrea Ferroglia$^{b,c}$,  Ben D. Pecjak$^d$, and Zhibai Zhang$^{b,c}$}

    \vspace{2ex}
  
    \textsl{${}^a$Paul Scherrer Institut,\\ CH-5232 Villigen PSI, Switzerland
      \\[0.3cm]
      ${}^b$Physics Department, New York City College of Technology,\\ The City University of New York
      \\ Brooklyn, NY 11201, USA
      \\[0.3cm]
${}^c$The Graduate School and University Center,\\ The City University of New York\\
New York NY 10016, USA
\\[0.3cm]
${}^d$Institute for Particle Physics Phenomenology,\\
University of Durham\\
DH1 3LE Durham, UK}
\end{center}

  \vspace{4ex}

  \begin{abstract}

    We derive the hard functions for all $2 \to 2$ processes in
    massless QCD up to next-to-next-to-leading order (NNLO) in the
    strong coupling constant.  By employing the known one- and
    two-loop helicity amplitudes for these processes, we obtain
    analytic expressions for the ultraviolet and infrared finite,
    minimally subtracted hard functions, which are matrices in color
    space.  These hard functions will be useful in carrying out higher-order
    resummations in processes such as dijet and highly
    energetic top-quark pair production by means of soft-collinear
    effective theory methods.

\end{abstract}

\end{titlepage}

\section{Introduction}
\label{sec:intro}

Some of the most fundamental processes at hadron colliders such as the
LHC are mediated at leading order (LO) in perturbation theory by
$2\to2$ scattering processes of colored particles -- two prime
examples within the Standard Model are dijet and top-quark pair
production.  Fixed-order perturbation theory provides an obvious and
conceptually straightforward framework in which to calculate
higher-order QCD corrections to the total and differential cross
sections for such processes, but it is often interesting or even
necessary to supplement the fixed-order calculations with certain
classes of logarithmic corrections to all orders in perturbation
theory. 

The factorization formulas underlying such resummations depend on the
way in which the observable is sensitive to soft and collinear
emissions, and are thus in general different for each particular
differential cross section. Concrete examples are threshold
resummation for inclusive jet production \cite{Kidonakis:1998bk,
  Kidonakis:2000gi, Kumar:2013hia}, highly boosted top-quark pair
production \cite{Ferroglia:2012ku, Ferroglia:2013awa}, and inclusive
hadroproduction \cite{Catani:2013vaa}, as well as the resummations
used for dijet event shapes in \cite{Banfi:2010xy}. 
Relatively recently, the factorization of processes with two (or more) jets was also studied by means of  Soft Collinear Effective Theory (SCET) methods
\cite{Bauer:2008jx,Bauer:2010vu}. 
 However, a common
ingredient to all resummations for $2\to 2$ processes are so-called
``hard functions'', which account for virtual corrections to the
underlying Born amplitudes.  There is thus one such hard function for
each possible $2\to 2$ partonic process involving massless quarks and gluons, although
all can be derived from those for $q\bar{q}\to Q\bar{Q}$, $qg\to qg$,
and $gg\to gg$ scattering, where $q$ and $Q$ are distinct 
quarks.  The hard functions are related to the interference of vector
components of color-decomposed helicity amplitudes, and are matrices
of varying dimension for the different partonic scattering processes
-- $2 \times 2$ for four-quark processes, $3 \times 3$ for $qg\to qg$
processes, and $9 \times 9$ for the four-gluon process.  The
next-to-leading order (NLO) hard functions for such QCD processes were
extracted in \cite{Kelley:2010fn}; these are a necessary ingredient for the resummation of any dijet hadronic process up to
next-to-next-to-leading logarithmic (NNLL) accuracy.

The goal of the current work is to build on previous results by
presenting the complete set of next-to-next-to-leading order (NNLO)
hard functions.   The main building blocks needed in this task are the
NNLO UV-renormalized, color-decomposed helicity amplitudes for $2\to2$
massless QCD processes calculated in
\cite{Glover:2003cm,Glover:2004si, Bern:2003ck, DeFreitas:2004tk,Bern:2002tk}. We turn these computations into
results for the hard functions by performing an IR renormalization
procedure on the color decomposed amplitudes, and then constructing
the spin-averaged hard matrices by interfering all possible combinations
of the fully renormalized color-decomposed amplitudes. Needless to
say, the final results are quite lengthy, and are therefore given in
electronic form with the arXiv submission of this work.  To facilitate
use by other groups, we also provide a {\tt Mathematica} interface to
the results.  

The hard functions we calculate in the present work are a necessary ingredient for pushing any resummed
calculation of a dijet observable at hadron colliders to next-to-next-to-next-to-leading logarithmic (NNNLL)
order, as they provide the boundary terms for the
renormalization-group evolution equations to that order.  However,
especially in cases where NNLO results are known, it is a frequent
practice to include these boundary terms on top of NNLL resummations
to achieve ``NNLL$'$+NNLO'' accuracy\footnote{Here we use the
  nomenclature of, e.g., \cite{Stewart:2013faa}.}, even in the absence
of the three-loop non-cusp and four-loop cusp anomalous dimensions
needed for a full NNNLL resummation.  We thus anticipate that the
results collected here will be useful for practitioners of
higher-order resummation in the near and distant future.

The organization of the paper is as follows: we describe our calculational procedure in Section~\ref{sec:calcproc},
give results in Section~\ref{sec:results}, and conclude in Section~\ref{sec:conclusions}.

\section{Hard functions to NNLO: calculational procedure}
\label{sec:calcproc}

The goal of this paper is to obtain the NNLO hard functions for all
scattering processes with two incoming and two outgoing partons in
massless QCD. These processes can be classified into groups containing
four quarks, two quarks and two gluons, and four gluons.  Including
all possible crossings, the four-quark processes are
\begin{align}
q(p_1) + \bar{q}(p_2) &\longrightarrow Q(p_3) + \bar{Q}(p_4) \, , \label{eq:qQs}  \\
q(p_1) + \bar{Q}(p_2) &\longrightarrow q(p_3) + \bar{Q}(p_4) \, ,  \label{eq:qQt} \\
q(p_1) + Q(p_2) &\longrightarrow q(p_3) + Q(p_4) \, , \label{eq:qQtp}   \\
q(p_1) + Q(p_2) &\longrightarrow Q(p_3) + q(p_4) \, ,  \label{eq:qQu}   \\
q(p_1) + \bar{q}(p_2) &\longrightarrow q(p_3) + \bar{q}(p_4) \, ,  \label{eq:qqst}  \\
q(p_1) + q(p_2) &\longrightarrow q(p_3) + q(p_4) \label{eq:qqutp} \, ,  
\end{align}
where $q$ and $Q$ indicate quarks of different flavors. For scattering 
processes involving two quarks and two gluons, we focus on the following
three possibilities:
\begin{align}
g(p_1) + g(p_2) &\longrightarrow q(p_3)  + \bar{q}(p_4) \, , \label{eq:qgs} \\
q(p_1) + g(p_2) &\longrightarrow q(p_3)  + g(p_4) \, ,   \label{eq:qgt}\\
q(p_1) + g(p_2) &\longrightarrow g(p_3)  + q(p_4) \, . \label{eq:qgu}
\end{align}
There is also a $q\bar q \to gg$ process, but with our 
definitions its hard function is the same as (\ref{eq:qgs}) up to an overall factor $(N^2-1)^2/N^2$ (where $N=3$ is the number of colors) which accounts for the  color average over the incoming quarks rather than the incoming gluons. Furthermore, one needs to consider processes
analogous to (\ref{eq:qgt}, \ref{eq:qgu}) containing antiquarks, for
which the hard functions are also identical to the corresponding processes involving quarks.  We therefore omit these from
the discussion.  Finally, we consider the four-gluon scattering
process
\begin{align}
g(p_1) + g(p_2) \longrightarrow g(p_3)  + g(p_4) \, . \label{eq:gggg} 
\end{align}

Here and in the following, we associate to the particle carrying
momentum $p_i$ a helicity index $\lambda_i \in \{+,-\}$ and a color index
$a_i$, which is understood to be in the fundamental representation of
SU$(3)$ if the particle is a quark, and in the adjoint representation
if the particle is a gluon.  For use later on, we introduce the
invariants
\begin{align}
s = (p_1+p_2)^2 \, , \qquad t = (p_1-p_3)^2 \, ,
\quad u = (p_1-p_4)^2 \, , \quad \text{and} \quad  r  = -t/s \, .
\end{align}
Momentum conservation implies that the Mandelstam variables satisfy
$s+t+u=0$,  which we will use to write our end results 
as functions of $s$, the 't Hooft scale $\mu$, and the dimensionless ratio $r$.

A unique hard function is associated to each of the processes listed
above.  These can all be extracted using the two-loop helicity
amplitudes calculated in \cite{Glover:2003cm,
  Glover:2004si,Bern:2003ck, DeFreitas:2004tk,Bern:2002tk}. To
describe the calculational procedure that goes into doing this, we
first introduce some aspects of the color-space formalism of
\cite{Catani:1996jh}, which allows us to treat the different cases
with a uniform notation.  In this formalism the UV-renormalized
helicity amplitudes are considered vectors in color space,
whose perturbative expansions we define as
\begin{align}
  |{\mathcal
    M_h}(\epsilon,r,s)\rangle & =
  4 \pi \alpha_s \left[|{\mathcal M}_h^{(0)}\rangle + \frac{\alpha_s}{2
      \pi} |{\mathcal
      M}_h^{(1)}\rangle +
    \left( \frac{\alpha_s}{2 \pi}\right)^2 |{\mathcal
      M}_h^{(2)}\rangle +
    {\mathcal O}{(\alpha_s^3)} \right] \,  \label{eq:Mexp} \\
   & = 4 \pi \alpha_s \left[|\hat{{\mathcal M}}_h^{(0)}\rangle + \frac{\alpha_s}{4
      \pi} |\hat{{\mathcal   M}}_h^{(1)}\rangle +
    \left( \frac{\alpha_s}{4 \pi}\right)^2 |\hat{{\mathcal
      M}}_h^{(2)}\rangle +
    {\mathcal O}{(\alpha_s^3)} \right] \, .  \label{eq:Mexp4pi}
\end{align}
Here $\epsilon = (4-d)/2$ is the dimensional regulator and the subscript
$h=(\lambda_1,\lambda_2,\lambda_3,\lambda_4)$ labels the helicity
amplitudes. Moreover, we have suppressed the arguments of the
expansion coefficients on the right-hand side.\footnote{These coefficients depend on
  on $r,s$, and the renormalization scale $\mu_r$, although the
  all-order amplitude on the left-hand side is independent of
  $\mu_r$.}  Finally, in order to follow SCET conventions for the perturbative 
  expansions of the hard functions in powers of $\alpha_s/4\pi$ below, 
  we have defined a set of coefficients $|\hat{{\mathcal M}}_h^{(L)}\rangle \equiv 2^L |{\mathcal M}_h^{(L)}\rangle.$

The amplitudes can be further decomposed in a
particular color-space basis as
\begin{equation}
|{\mathcal M}_h^{(L)} \rangle = \sum_{I=1}^n {\mathcal M}^{(L)}_{hI} | {\mathcal C}_I \rangle \, , 
\label{eq:Mincol}
\end{equation}
where $| {\mathcal C}_I \rangle $ are basis vectors. The basis
includes two vectors in processes involving four quarks and three
vectors in processes involving two quarks and two gluons. For the
four-gluon scattering process we will use the redundant basis
involving nine vectors employed in \cite{Bern:2002tk}. 

The helicity amplitudes contain IR poles in the dimensional
regulator $\epsilon$.  We can subtract these poles in the $\msbar$
scheme using the renormalization procedure described in
\cite{Becher:2009cu, Becher:2009qa}.
We thus define renormalized amplitudes according to
\begin{equation}
\label{eq:IRsub}
|{\mathcal M}_h^{\text{ren}} \left(r,s,\mu \right)\rangle  = \lim_{\epsilon \to 0} \bm{Z}^{-1}\left(\epsilon,r,s , \mu \right)
|{\mathcal M}_h\left(\epsilon, r,s\right) \rangle\, .
\end{equation}
The exact form of the renormalization factor $\bm{Z}$ was determined
up to two-loops by means of SCET methods in \cite{Becher:2009cu,
  Becher:2009qa}; for now
we just note that it is the same for all helicity amplitudes.  For reasons that will become
apparent later on, we define the perturbative expansion of the renormalization factor as
(with a slight abuse of notation inherited from
\cite{Becher:2013vva})
\begin{align}
\bm{Z}^{-1} \left(\epsilon, r, s, \mu \right) = \bm{1} + \frac{\alpha_s}{2 \pi} \bm{Z}^{(1)} \left(\epsilon\right)
+ \left(\frac{\alpha_s}{2 \pi} \right)^2 \bm{Z}^{(2)} \left(\epsilon \right) + {\mathcal O}\left( \alpha_s^3\right),
\end{align}
where $\bm{Z}^{(L)}(\epsilon)\equiv \bm{Z}^{(L)}(\epsilon,r,s,\mu)$.
We can then evaluate (\ref{eq:IRsub}) order-by-order in perturbation
theory, defining renormalized amplitudes and expansion coefficients
analogous to (\ref{eq:Mexp}) and (\ref{eq:Mincol}).

With this notation it is now a simple matter to write expressions for
the hard functions to NNLO.   We first define expansion coefficients
through
\begin{align}
\bm{H}(r,s,\mu) = 
\frac{16 \pi^2\alpha_s^2}{{\cal N}_R} \left[\bm{H}^{(0)}  + \frac{\alpha_s}{4 \pi} \bm{H}^{(1)}  + \left(\frac{\alpha_s}{4 \pi} \right)^2
 \bm{H}^{(2)}  + {\mathcal O}\left(\alpha_s^3 \right)  \right] \, ,
\end{align}
where $\bm{H}^{(L)}(r,s,\mu)\equiv \bm{H}^{(L)}$.   
The factor $\mathcal{N}_R$ takes into account the channel-dependent 
factors related to averaging over initial state colors; it is defined
as $\mathcal{N}_R = N^2$ for initial states with two quarks,  
$\mathcal{N}_R = N (N^2-1)$ for quark-gluon initial states, and 
$\mathcal{N}_R= (N^2-1)^2$ for initial states with two gluons.
In terms of the color-decomposed, IR and UV
renormalized helicity amplitudes perturbatively expanded as in (\ref{eq:Mexp4pi}), the hard function matrix elements
read 
\begin{align}
H^{(0)}_{IJ}  &=\frac{1}{4}\sum_h \left(\hat{{\mathcal M}}^{(0)}_{hI} \right)^* \hat{{\mathcal M}}^{(0)}_{hJ} \, ,\nn \\
H^{(1)}_{IJ}  &=\frac{1}{4}\sum_h  \left[\left(\hat{{\mathcal M}}^{(0)}_{hI} \right)^* \hat{{\mathcal M}}^{(1)}_{hJ} +\left(\hat{{\mathcal M}}^{(1)}_{hI} \right)^* \hat{{\mathcal M}}^{(0)}_{hJ}   \right]\, ,\nn \\
H^{(2)}_{IJ}  &=\frac{1}{4}  \sum_h \left[
\left(\hat{{\mathcal M}}^{(1)}_{hI} \right)^* \hat{{\mathcal M}}^{(1)}_{hJ}    + \left(\hat{{\mathcal M}}^{(0)}_{hI} \right)^* \hat{{\mathcal M}}^{(2)}_{hJ} +\left(\hat{{\mathcal M}}^{(2)}_{hI} \right)^* \hat{{\mathcal M}}^{(0)}_{hJ}  \right]\, . \label{eq:Horder}
\end{align}
The factor $1/4$ in~(\ref{eq:Horder}) is related to the average
over the spin of the two incoming partons.  The normalization of the expansion coefficients 
above (but not that of the hard function itself) then coincides with the $m_t \to 0$ limit of the 
corresponding results for the production of massive top pairs \cite{Ahrens:2010zv}.
The hard functions are $2 \times 2$ matrices for scattering processes
involving four quarks and $3 \times 3$ matrices for processes involving
two quarks and two gluon. The hard functions for the process involving
four gluons are $9 \times 9$ matrices with our choice of color
basis. All of the matrices are Hermitian.

With our definitions, the hard function is related to the square of
the renormalized amplitude as
\begin{align}
\frac{1}{4 {\cal N}_R} \sum_h \langle {\cal M}_h^\text{ren}\left(r,s,\mu \right) | {\cal M}_h^\text{ren} \left(r,s,\mu \right) \rangle = \mbox{Tr}\left[ \bm{H}\left(r,s,\mu \right)
  \bm{\tilde{s}}^{(0)}\right] \,. \label{eq:TrHsvM2}
\end{align}
The matrix $\bm{\tilde{s}}^{(0)}$ is a ``tree-level soft function",
whose elements are defined as 
\begin{equation}
\label{eq:stree}
\tilde{s}^{(0)}_{IJ} =  \langle {\mathcal C}_I |  {\mathcal C}_J \rangle \, .
\end{equation}
The color bases for the various processes we
consider are specified in the next section, along with the
basis-dependent results for the  soft functions (\ref{eq:stree}).
Furthermore, the hard function is related to the $L$-loop corrections to
the double differential partonic cross section in $s$ and $r$ by
\begin{align}
\frac{d^2 \hat{\sigma}^{(L)}}{d s d r} = 
\frac{\alpha_s^2}{\mathcal{N}_R} \left(\frac{\alpha_s}{  4 \pi}\right)^{L}
\frac{\pi}{s^2} \mbox{Tr}\left[ \bm{H}^{(L)} \left(r,s,\mu \right)
  \bm{\tilde{s}}^{(0)}\right] \, .
\end{align}

With this conceptual framework in place, we now address the more
practical issue of how to extract the hard functions from the NNLO
helicity amplitudes calculated in \cite{Glover:2003cm,Glover:2004si,
  Bern:2003ck, DeFreitas:2004tk,Bern:2002tk}. In all cases, we have
used the helicity amplitudes evaluated in the 't~Hooft-Veltman (HV)
scheme.
The most straightforward way to use the information in those papers
would be to construct the UV-renormalized, color decomposed helicity
amplitudes, perform the IR renormalization procedure (\ref{eq:IRsub}),
and then evaluate the matrix elements (\ref{eq:Horder}).  We have
indeed used this straightforward (and tedious) method in obtaining our
results.

A slightly more streamlined method, detailed recently in
\cite{Becher:2013vva}, uses that \cite{Glover:2003cm,
  Glover:2004si,Bern:2003ck, DeFreitas:2004tk,Bern:2002tk} do in fact
quote results for IR finite amplitudes, but in different
renormalization schemes based on the structure of IR poles written
down in \cite{Catani:1998bh}.  Therefore, constructing the $\msbar$
subtracted amplitudes (\ref{eq:IRsub}) from those works is just a
matter of switching between renormalization schemes.  To understand
how to perform this switch, we first consider the typical split of
UV-renormalized helicity amplitudes into pole and finite remainder
terms used in \cite{Glover:2003cm, Glover:2004si,Bern:2003ck,
  DeFreitas:2004tk,Bern:2002tk}.  As a concrete example, the one-loop
helicity amplitudes calculated in \cite{Glover:2004si} are written as
(see~(4.10) of that work)
\begin{equation}
|{\mathcal M}^{(1)}_{h} \rangle = \bm{I}^{(1)}(\epsilon) 
|{\mathcal M}^{(0)}_{h} \rangle  + 
|{\mathcal M}^{(1),\text{fin}}_{h} \rangle \, , \label{eq:ampex}
\end{equation}
while the two-loop helicity amplitudes are written as (see~(4.11)
of that work)
\begin{equation}
\label{eq:ampex2}
 |{\mathcal M}^{(2)}_{h}
\rangle =
\bm{I}^{(2)}(\epsilon) |{\mathcal M}^{(0)}_{h}
\rangle +
\bm{I}^{(1)}(\epsilon) |{\mathcal M}^{(1)}_{h} 
\rangle+ |{\mathcal
  M}^{(2),\text{fin}}_{h}
\rangle \, .  
\end{equation}
The IR poles are contained in the color-space operator $\bm{I}$
\cite{Catani:1998bh}, whose perturbative expansion is defined as
\begin{align}
\bm{I}(\epsilon,r,s,\mu) = \bm{1}+\frac{\alpha_s}{2\pi}\bm{I}^{(1)}(\epsilon)+
\left(\frac{\alpha_s}{2\pi}\right)^2 \bm{I}^{(2)}(\epsilon) + {\mathcal O}\left( \alpha_s^3\right) \,
\end{align}
where $\bm{I}^{(L)}(\epsilon) \equiv \bm{I}^{(L)}(\epsilon,r,s,\mu)$.
This object is analogous, but not identical, to the renormalization
factor $\bm{Z}$ in (\ref{eq:IRsub}).  The difference is that $\bm{Z}$
contains only pole terms while $\bm{I}$ contains both pole terms and
some finite terms. Moreover, the $1/\epsilon$ pole term and finite
parts of the two-loop coefficient $\bm{I}^{(2)}$ were not fully
specified in \cite{Catani:1998bh}, but are instead parameterized in a
function $\bm{H}^{(2)}_{R.S.}$ defined in equation (19) of that work.  The
authors of \cite{Glover:2003cm, Glover:2004si,Bern:2003ck,
  DeFreitas:2004tk,Bern:2002tk} provide explicit expressions for this
function in their calculations, but in such a way that the finite parts of
$\bm{I}^{(2)}$ are not the same in each paper.  For these reasons, the
finite remainders quoted in \cite{Glover:2003cm,
  Glover:2004si,Bern:2003ck, DeFreitas:2004tk,Bern:2002tk} differ from
the $\msbar$ renormalized amplitudes (\ref{eq:IRsub}). Instead, they
can be viewed as renormalized amplitudes in a scheme defined by
equations (\ref{eq:ampex}) and (\ref{eq:ampex2}) above, which differs from
calculation to calculation according to the exact choice of
$\bm{H}^{(2)}_{R.S.}$.
To convert these finite-remainders to the $\msbar$ scheme, one can insert
(\ref{eq:ampex}), (\ref{eq:ampex2}) into (\ref{eq:IRsub}) to find
\begin{align}
|{\mathcal M}_h^{(1),\text{ren}} \rangle  &= |{\mathcal M}_h^{(1),\text{fin}} \rangle + \left(\bm{I}^{(1)}(\epsilon) + \bm{Z}^{(1)}(\epsilon) \right) |{\mathcal M}_h^{(0)} \rangle \, , \nonumber \\
|{\mathcal M}_h^{(2),\text{ren}} \rangle  &=|{\mathcal M}_h^{(2),\text{fin}} \rangle + \left(\bm{I}^{(1)}(\epsilon) + \bm{Z}^{(1)}(\epsilon) \right) |{\mathcal M}_h^{(1),\text{fin}} \rangle \nn \\
& + \left[\bm{I}^{(2)}(\epsilon) + \left( \bm{I}^{(1)}(\epsilon) + \bm{Z}^{(1)}(\epsilon)\right) \bm{I}^{(1)}(\epsilon) +
\bm{Z}^{(2)}(\epsilon) \right] |{\mathcal M}_h^{(0)} \rangle \, . \label{eq:finamp}
\end{align}
One can then recast this equation into an explicitly IR finite form.
As explained above, the result depends on the choice of the single
pole term in $\bm{I}^{(2)}$.  In the case where this term is identical
to that in $\bm{Z}$, i.e. adds no extra finite parts to
$\bm{I}^{(2)}$,
one can write the result in terms of $\epsilon$ independent operators
$\bm{{\mathcal C}}_i$ ($i = 0,1$) and various known anomalous
dimensions as in \cite{Becher:2013vva}
\begin{align}
\bm{I}^{(1)}(\epsilon) + \bm{Z}^{(1)}(\epsilon) & = \bm{{\mathcal C}}_0 \, , \nn \\
\bm{I}^{(2)}(\epsilon) +\left( \bm{I}^{(1)}(\epsilon) + \bm{Z}^{(1)}(\epsilon)\right) \bm{I}^{(1)}(\epsilon) +
\bm{Z}^{(2)}(\epsilon) & = 
\frac{1}{2} \bm{{\mathcal C}}^2_0 +\frac{\gamma_1^{\text{cusp}}}{8} \left( \bm{{\mathcal C}}_0 + \frac{\pi^2}{128} \Gamma'_0\right) \nn \\
&+ \frac{\beta_0}{2} \left(\bm{{\mathcal C}}_1 + \frac{\pi^2}{32} \bm{\Gamma}_0 + \frac{7 \zeta_3}{96} \Gamma'_0 \right)
-\frac{1}{8} \left[ \bm{\Gamma}_0, \bm{{\mathcal C}}_1\right] \, , \label{eq:Cs}
\end{align}
where
\begin{align}
\bm{\Gamma}_0 &= \sum_{(i,j)} \frac{\bm{T}_i \cdot \bm{T}_j}{2} \gamma_0^{\text{cusp}} \ln{\frac{\mu^2}{-s_{ij}}}
+ \sum_i \gamma_0^i \, , \nn \\ 
\bm{{\mathcal C}}_0 
&= \sum_{(i,j)} \frac{\bm{T}_i \cdot \bm{T}_j}{16} \left[ \gamma_0^{\text{cusp}} \ln^2{\frac{\mu^2}{-s_{ij}}} 
- 4 \frac{\gamma_0^i}{C_i} \ln{\frac{\mu^2}{-s_{ij}}} \right] - \frac{\pi^2}{96} \Gamma'_0 \, , \nn \\
\bm{{\mathcal C}}_1 
&= \sum_{(i,j)} \frac{\bm{T}_i \cdot \bm{T}_j}{48} \left[ \gamma_0^{\text{cusp}} \ln^3{\frac{\mu^2}{-s_{ij}}} 
- 6 \frac{\gamma_0^i}{C_i} \ln^2{\frac{\mu^2}{-s_{ij}}} \right] - \frac{\pi^2}{48} \bm{\Gamma}_0 \, 
- \frac{\zeta_3}{24} \Gamma'_0  \, . \label{eq:operators}
\end{align}
The sums run over the unordered tuples $(i,j)$ of distinct partons,
$\bm{T}_i$ is the color generator associated with the i-th parton in
the scattering amplitude, and $s_{ij} \equiv 2 \sigma_{ij} p_i \cdot
p_j + i0$, where the sign factor $\sigma_{ij} = +1$ if the two parton
momenta are both incoming or outgoing, and $\sigma_{ij} = -1$
otherwise.  The anomalous dimensions appearing in~(\ref{eq:Cs},\ref{eq:operators}) 
are collected in many places, for example in
Appendix A of \cite{Becher:2013vva}; an explicit expression for the
commutator $[\bm{\Gamma}_0,\bm{C}_1]$ in terms of $\bm{T}_i$,
logarithms and anomalous dimensions can be found in the same
appendix. We emphasize that in
\cite{Glover:2003cm,Glover:2004si,Bern:2003ck,
  DeFreitas:2004tk,Bern:2002tk} the single pole term in
$\bm{H}^{(2)}_{R.S.}$ is multiplied by factors of the form
$(-\mu^2/s_{ij})^{2 \epsilon}$, which yield additional finite
contributions to $\bm{I}^{(2)}$ and thus to the right-hand side of the
second line of (\ref{eq:Cs}) upon expansion in $\epsilon$.  We do not
list these explicitly, as they differ between
\cite{Glover:2003cm,Glover:2004si,Bern:2003ck,
  DeFreitas:2004tk,Bern:2002tk}, but we do take them into account when
extracting $\msbar$-renormalized helicity amplitudes from those
references using (\ref{eq:finamp}) and (\ref{eq:Cs}).  

We have calculated the $\msbar$-renormalized helicity amplitudes using
both methods described above, and checked that they agree.  We then
used these amplitudes to construct the hard functions through
(\ref{eq:Horder}).  We end this section by describing several
cross-checks we have performed on our channel and basis-dependent
results, which we give in the next section. First, for the $q\bar{q} \to Q
\bar{Q}$ and $g g \to q \bar{q}$ channels, we verified that the trace
of functions given in (\ref{eq:TrHsvM2}) is consistent with the NNLO
results derived in \cite{Ferroglia:2013zwa}. In turn, the results in
the latter reference were tested against the squared NLO and NNLO
matrix elements for the processes in~(\ref{eq:qQs},\ref{eq:qgs}),
which can be found in \cite{Anastasiou:2001sv,Anastasiou:2000kg,
  Anastasiou:2000mv}.  For the $gg\to gg$ channel, on the other hand,
we have checked (\ref{eq:TrHsvM2}) against the UV-renormalized squared
matrix elements given in \cite{Glover:2001af,
  Glover:2001rd}. In order to carry out this last comparison, it was necessary to renormalized away the IR poles from the squared amplitudes in  \cite{Glover:2001af,
    Glover:2001rd}, this was done by employing once more the IR renormalization method of \cite{Becher:2009cu, Becher:2009qa}.

Second, the hard functions for the channels in~(\ref{eq:qQt},\ref{eq:qQtp},\ref{eq:qQu}) were assembled not only
by starting from the appropriate amplitudes obtained from
\cite{Glover:2004si}, but also by applying crossing symmetries to the
amplitudes for the process in~(\ref{eq:qQs}). Similarly, the hard
functions for the processes in~(\ref{eq:qgt},\ref{eq:qgu}) were
also obtained a second time from the amplitudes for the process in~(\ref{eq:qgs}) by applying crossing symmetries.

Third, we have checked that the hard functions satisfy the 
renormalization-group equations implied by (\ref{eq:IRsub}).  These take the form 
\be
\frac{d}{d \ln \mu} \bm{H}(r,s,\mu) = \bm{\Gamma}(r,s,\mu)  \bm{H}(r,s,\mu) + \bm{H}(r,s,\mu) \bm{\Gamma}^\dagger(r,s,\mu) \, .
\label{eq:RGE}
\ee
The general form of the anomalous dimension operator $\bm{\Gamma}$ in
the case of massless scattering amplitudes is exactly known up to two loops
\cite{Becher:2009cu, Becher:2009qa}. In those papers it is also conjectured
that $\bm{\Gamma}$ involves only color dipoles at all orders.
The $n$-th order term in the expansion of $\bm{\Gamma}$ can be obtained by replacing $\gamma^a_0 \to \gamma^a_n$ ($a \in \{\mbox{cusp}, i\}$) in~(\ref{eq:operators}). We give explicit, channel and basis dependent
results for the anomalous dimension in the next section. 

As a byproduct of our calculation we also evaluated the NLO hard
functions, which were previously calculated in \cite{Kelley:2010fn}. We
find agreement with the results in that work, after we account
for differences in notation.\footnote{Equation~(55) in \cite{Kelley:2010fn}
  defines a real hard function in the channels with two quarks and two
  gluons, while we define Hermitian hard functions with complex off
  diagonal terms. Furthermore, one needs to be careful when applying
  the crossing relations listed in Table~2 of \cite{Kelley:2010fn} in
  cases where fermions are switched between the initial and final
  state, as this necessitates extra minus signs. Finally,
  there is a minor typo in Table~5 of \cite{Kelley:2010fn}, which is
  related to the four-gluon channel: the color factors listed in the
  next to the last column of that table apply to the helicities
  labeled 7 and 8, while the ones listed in the last column of the
  table apply to the helicities labeled by the numbers 9-16.}

\section{Hard functions to NNLO: results}
\label{sec:results}

We now present our results for the hard functions. We split the
discussion into three subsections for the four-quark, two quark plus
two gluon, and four-gluon scattering, which in turn are subdivided according
to momentum crossings. In each case we define the
channel-dependent color basis in which the hard function is
calculated, and give analytic results for the tree-level hard and soft
functions as well as the anomalous dimension $\bm{\Gamma}$ in that
basis.  The color bases are defined by projections of the basis
vectors onto the arbitrary vector $| \{a\}\rangle \equiv | a_1, a_2,
a_3, a_4 \rangle $, where $a_i$ represents the color index of the
parton $i$ (which can be either in the fundamental or adjoint
representation, depending on the process).  These vectors satisfy the
relation
\begin{equation}
\langle \{a\}| \{b\} \rangle  = \delta_{a_1 b_1}  \delta_{a_2 b_2}  \delta_{a_3 b_3}  \delta_{a_4 b_4} \, .
\end{equation}
The action of the color operators $\bm{T}_i$ on the vectors $|
\{a\}\rangle$, which is needed to construct the basis-dependent
expressions, is discussed in many references, see for example Section
3.2 in \cite{Ahrens:2010zv}.

The main results of this work are the NNLO hard
functions obtained through the last line of~(\ref{eq:Horder}). The analytic results
for these functions would fill about 100 pages, were they printed out explicitly.
As by now customary in such situations, we instead include the results in
electronic format with the arXiv submission of this work. All of the
hard functions are stored in {\tt Mathematica} input files which can
be loaded in the accompanying {\tt Mathematica} notebook. In the
latter file, a simple function allows the user to obtain numerical
values for the hard functions for the processes listed in~(\ref{eq:qQs}-\ref{eq:gggg}) once the desired perturbative order
(LO, NLO, or NNLO) and the values of $r,s,\mu$ and $N_l$ (the number of fermions) are
specified. As a reference for other groups which might desire to carry
out this calculation, we give explicit numerical results for the
NLO and NNLO hard functions at a specific benchmark point in the subsections
that follow.  In all cases we use 
\begin{align}
\label{eq:benchmark}
N=3, \quad N_l = 5, \quad r = \frac{\sqrt{5} - 1}{2}, \quad \sqrt{s} = 2\mu .
\end{align}

\subsection{Four-quark scattering}
Here we summarize results for the four-quark scattering
processes in~(\ref{eq:qQs}--\ref{eq:qqutp}).  In all cases
we use singlet-octet type color bases defined below, for which the tree-level 
soft function is\footnote{The soft function in~(\ref{eq:s0qQ}) differs 
by an overall factor $N$ from the one some of us employed in previous work involving four quark partonic processes  (see for example \cite{Ahrens:2010zv}). This is due to the definition of ${\bm \tilde{s}}^{(0)}$ in~(\ref{eq:stree}), which differs slightly from the one employed in previous papers. Analogous considerations apply to the soft function for the channels involving two gluons, which can be found in~(\ref{eq:s0qg}).
}
\begin{align}
\bm{\tilde{s}}^{(0)} = \left( 
\begin{array}{cc}
N^2 &  0\\ 
 0 & \frac{C_F N}{2}
\end{array} 
\right) \, , \label{eq:s0qQ}
\end{align}
where $C_F = (N^2-1)/(2N)$.  Channel dependent results for the hard functions and 
anomalous dimensions are gathered in the subsections below.

\boldmath
\subsubsection{$q(p_1) + \bar{q}(p_2) \to Q(p_3) + \bar{Q}(p_4)$}
\unboldmath

The color basis which we employ to describe the four-quark process in~(\ref{eq:qQs}) is 
\be
{\mathcal C}_1 \equiv  \langle \{a\}| {\mathcal C}_1 \rangle  = \delta_{a_1 a_2} \delta_{a_3 a_4}  \, , \qquad
{\mathcal C}_2 \equiv  \langle \{a\}| {\mathcal C}_2 \rangle  = t^c_{a_2 a_1} t^c_{a_3 a_4} \, .\label{eq:colqQs}
\ee
with this choice, the tree-level hard function is 
\be
\bm{H}^{(0)} = \left( 1 - 2 r + 2 r^2\right)\left( \begin{array}{cc}
0 & 0  \\ 
0 & 2
\end{array} \right) \, . \label{eq:LOqqbQQb}
\ee
We also provide numerical values of the NLO and NNLO
hard functions at the  benchmark point (\ref{eq:benchmark}):
\begin{align}
\bm{H}^{(1)} = \left(\begin{array}{cc}
0 &  -0.139210 - i\, 0.192224 \\ 
-0.139210 + i\, 0.192224  & 2.51146
\end{array}  \right)\, ,
\end{align}
\begin{align}
\bm{H}^{(2)} = \left(\begin{array}{cc}
7.16744 & -22.1589 - i\, 70.2433\\ 
 -22.1589 + i\, 70.2433 & 380.359
\end{array}  \right) \, .
\end{align}
The anomalous dimension $\bm{\Gamma}$ in this basis is 
\begin{align}
\label{eq:G125}
\bm{\Gamma} &= \left[2 C_F \gamma_{\text{cusp}} \left(\alpha_s \right) \left(\ln{\frac{s}{\mu^2}} -i \pi \right) + 4 \gamma^q\left(\alpha_s \right)\right] \bm{1} \nn \\
& +N \gamma_{\text{cusp}} \left(\alpha_s \right) \left( \ln{r} +i \pi \right) \left(
\begin{array}{cc}
0 & 0 \\ 
0 & 1
\end{array} 
\right) +
2 \gamma_{\text{cusp}} \left(\alpha_s \right) \ln\left({\frac{r}{1-r}}\right) \left(
\begin{array}{cc}
0 & \frac{C_F}{2 N} \\ 
1 & - \frac{1}{N} 
\end{array} 
\right) \, .
\end{align}
The anomalous dimensions $\gamma_{\text{cusp}}$ and $\gamma^q$ have an expansion in powers of $a \equiv \alpha_s/(4 \pi)$ of the form $\gamma = \sum_i a^i \gamma_i$. The coefficients of the expansions up to NNLO  are collected in many sources, for example Appendix A in \cite{Becher:2013vva}.

\boldmath
\subsubsection{$q(p_1) + \bar{Q}(p_2) \to q(p_3) + \bar{Q}(p_4)$ }
\unboldmath

The color basis which we employ to describe the four-quark process in~(\ref{eq:qQt}) is the same one introduced in 
~(\ref{eq:colqQs}). The tree-level hard matrix is 
\be
\bm{H}^{(0)} =  \frac{2 - 2 r +  r^2}{N^2 r^2}\left( \begin{array}{cc}
\frac{(N^2-1)^2}{2N^2} & {-\frac{N^2-1}{N}}  \\ 
{-\frac{N^2-1}{N}}  & 2
\end{array} \right) \, . \label{eq:LOqQbqQb}
\ee
The NLO and NNLO matrices at the benchmark point (\ref{eq:benchmark}) are 
\begin{align}
\bm{H}^{(1)} &= \left(\begin{array}{cc}
71.4641 & -47.5473 -i \,13.8859\\ 
  -47.5473 +i \,13.8859 &  31.1224
\end{array}  \right)\, , \nn \\
\bm{H}^{(2)} &= \left(\begin{array}{cc}
2231.12 & -1575.43 - i \, 608.692\\ 
-1575.43 + i \, 608.692 & 1325.99
\end{array}  \right)\, ,
\end{align}
and the anomalous dimension $\bm{\Gamma}$ in this basis is the same as (\ref{eq:G125}).

\boldmath
\subsubsection{$q(p_1) + Q(p_2) \to q(p_3) + Q(p_4)$}
\unboldmath

The color basis which we employ to describe the four-quark process in~(\ref{eq:qQtp}) is 
\be
{\mathcal C}_1 \equiv  \langle \{a\}| {\mathcal C}_1 \rangle  = \delta_{a_3 a_2} \delta_{a_4 a_1}  \, , \qquad
{\mathcal C}_2 \equiv  \langle \{a\}| {\mathcal C}_2 \rangle  = t^c_{a_3 a_2} t^c_{a_4 a_1} \, .\label{eq:colqQtp}
\ee
In this channel, the tree-level hard matrix is identical to the one in~(\ref{eq:LOqQbqQb}).  The NLO and NNLO matrices at the benchmark point (\ref{eq:benchmark})
are 
\begin{align}
\bm{H}^{(1)} &= \left(\begin{array}{cc}
58.9143 & -50.2365 + i\, 13.8859\\ 
 -50.2365 - i\, 13.8859 &  42.2155
\end{array}  \right)\, , \nn \\
\bm{H}^{(2)} &= \left(\begin{array}{cc}
2083.37 & -1350.52 +i\, 209.622\\ 
-1350.52 -i\, 209.622 & 1071.72
\end{array}  \right)\, ,
\end{align}
and the anomalous dimension $\bm{\Gamma}$ in this basis is
\begin{align}
\label{eq:G346}
\bm{\Gamma} &= \left[2 C_F \gamma_{\text{cusp}} \left(\alpha_s \right) \left(\ln{\frac{s}{\mu^2}} -i \pi \right) + 4 \gamma^q\left(\alpha_s \right)\right] \bm{1} \nn \\
&+\gamma_{\text{cusp}} \left(\alpha_s \right) \left( \ln{r} +i \pi \right) \left(
\begin{array}{cc}
2 C_F & \frac{C_F}{N} \\ 
2 & \frac{N^2-3}{N}
\end{array} 
\right) +
\gamma_{\text{cusp}} \left(\alpha_s \right) \ln\left({\frac{r}{1-r}}\right) \left(
\begin{array}{cc}
- 2 C_F & 0 \\ 
0 &  \frac{1}{N} 
\end{array} 
\right) \, .
\end{align}

\boldmath
\subsubsection{$q(p_1) + Q(p_2) \to Q(p_3) + q(p_4)$}
\unboldmath

The color basis employed for the process in~(\ref{eq:qQu}) is the one we wrote in~(\ref{eq:colqQtp}).
The LO hard function in this channel is
\be
\bm{H}^{(0)} =  \left( 1 - \frac{2}{1-r}  + \frac{2}{(1-r)^2}\right)\left( \begin{array}{cc}
0 & 0  \\ 
0 & 2
\end{array} \right) \, .
\ee
The NLO and NNLO matrices at the benchmark point (\ref{eq:benchmark}) are
\begin{align}
\bm{H}^{(1)} &= \left(\begin{array}{cc}
 0 & -14.8225 - i\, 15.5573\\ 
-14.8225 + i\, 15.5573 &  1086.44
\end{array}  \right)\, , \nn \\
\bm{H}^{(2)} &= \left(\begin{array}{cc}
  36.6860 & 538.071 + i \, 2019.16\\ 
538.071 - i \, 2019.16  & 43819.6
\end{array}  \right)\, . 
\end{align}
The anomalous dimension $\bm{\Gamma}$ in this basis is the same as (\ref{eq:G346}).
\boldmath
\subsubsection{$q(p_1) + \bar{q}(p_2) \to q(p_3) + \bar{q}(p_4)$}
\unboldmath

We consider here the scattering process in~(\ref{eq:qqst}), where
we employ the color basis in~(\ref{eq:colqQs}).  The  tree-level
hard matrix is in this case given by
\begin{align}
\bm{H}^{(0)} = \frac{(N^2-1)}{N^3 r^2}\left( \begin{array}{cc}
\frac{(N^2-1)}{2 N} \left[2 - r (2-r)\right]  & - r \left(N
   (r-1)^2+r-2\right)-2 \\ 
-r \left(N
   (r-1)^2+r-2\right)-2 & 
\frac{N\left\{
2 r \left[N^2 r \left(2 (r-1)
r+1\right)+2 N
(r-1)^2+r-2\right]+4\right\}}{N^2-1}
\end{array} \right) \, .
\end{align}
The NLO and NNLO matrices at the benchmark point (\ref{eq:benchmark}) are
\begin{align}
\bm{H}^{(1)} &= \left(\begin{array}{cc}
71.9130 & -54.4697 - i\,14.8627\\ 
-54.4697 + i\,14.8627 &  45.2964
\end{array}  \right)\, , \nn \\
\bm{H}^{(2)} &= \left(\begin{array}{cc}
 2214.83  & -1745.69 - i\, 590.364\\ 
-1745.69 + i\, 590.364 & 2303.46
\end{array}  \right)\, ,
\end{align}
and the anomalous dimension $\bm{\Gamma}$ in this basis is the same as (\ref{eq:G125}).

\boldmath
\subsubsection{$q(p_1) + q(p_2) \to q(p_3) + q(p_4)$}
\unboldmath

The color basis which we employ in the process in~(\ref{eq:qqutp}) is the one in~(\ref{eq:colqQtp}). The tree-level hard matrix is
\begin{align}
\bm{H}^{(0)} = \frac{(N^2-1)}{N^3 r^2}\left( \begin{array}{cc}
\frac{(N^2-1)}{2 N} \left[2 - r (2-r)\right]  &  
\frac{2-r \left[N+(r-3) r+4\right]}{r-1}
 \\ \frac{2-r \left[N+(r-3) r+4\right]}{r-1}
 & 
\frac{N\left\{
2 r \left[ N^2
   \left(r^3+r\right)+2
N (r-1)+(r-2) \left((r-2)
   r+3\right)\right]+4
\right\}}{\left(N^2-1\right) \left(1-r\right)^2}
\end{array} \right) \, .
\end{align}
The NLO and NNLO matrices at the benchmark point (\ref{eq:benchmark}) are 
\begin{align}
\bm{H}^{(1)} &= \left(\begin{array}{cc}
50.4474 & 126.555 + i \, 2.12346\\ 
126.555 - i \, 2.12346 &  797.387
\end{array}  \right)\, , \nn \\
\bm{H}^{(2)} &= \left(\begin{array}{cc}
2106.09 & 5787.25 +i\, 2270.79\\ 
5787.25 -i\, 2270.79 & 35362.4
\end{array}  \right)\, ,
\end{align}
and the anomalous dimension $\bm{\Gamma}$ in this basis is the same as (\ref{eq:G346}).
\boldmath
\subsection{Two-quark two-gluon scattering}
\unboldmath 

We now turn to the two-quark two-gluon processes in
~(\ref{eq:qgs}--\ref{eq:qgu}).  As with the four-quark processes,
we choose color bases for the three processes such that the tree-level
soft function is the same for each, and reads 
\begin{align}
\bm{\tilde{s}}^{(0)} = V\left( \begin{array}{ccc}
N  & 0 & 0 \\ 
0 & \frac{N}{2}  & 0 \\ 
0 & 0  & \frac{N^2-4}{2 N} 
\end{array}  \right) \, , \label{eq:s0qg}
\end{align}
where we introduced the quantity $V \equiv N^2-1$.
Channel dependent
results for the hard functions and anomalous dimensions are gathered
in the subsections below.

\boldmath
\subsubsection{$g(p_1) + g(p_2) \to q(p_3)  + \bar{q}(p_4)$}
\unboldmath

The quark-antiquark pair production in the gluon fusion channel,~(\ref{eq:qgs}), is studied by employing the color basis
\begin{align}
{\mathcal C}_1 \equiv  \langle \{a\}| {\mathcal C}_1 \rangle  = \delta^{a_1 a_2} \delta_{a_3 a_4}  \, ,& \qquad&
{\mathcal C}_2 \equiv  \langle \{a\}| {\mathcal C}_2 \rangle  = i f^{a_1 a_2 c} t^c_{a_3 a_4 } \, , \nonumber \\
&{\mathcal C}_3 \equiv  \langle \{a\}| {\mathcal C}_3 \rangle  = d^{a_1 a_2 c} t^c_{a_3 a_4 } &\, . \label{eq:colggs}
\end{align}
With this basis, the tree-level hard matrix becomes
\begin{align}
\bm{H}^{(0)} = \left( 
\begin{array}{ccc}
\frac{1}{N^2} \left( \frac{1}{2 r} + \frac{1}{2(1-r)} -1\right) &  \frac{1}{N} \left( \frac{1}{2 r} - \frac{1}{2 (1-r)} + 2r -1\right)&  \frac{1}{N} \left(\frac{1}{2 r} + \frac{1}{2(1-r)} -1 \right) \\ 
\frac{1}{N} \left( \frac{1}{2 r} - \frac{1}{2 (1-r)} + 2r -1\right)& \frac{1}{2 r}+ \frac{1}{2 (1-r)}+ 4 r- 4 r^2 -3&   \frac{1}{2 r} - \frac{1}{2 (1-r)} + 2r -1 \\ 
\frac{1}{N} \left(\frac{1}{2 r} + \frac{1}{2(1-r)} -1 \right) & \frac{1}{2 r} - \frac{1}{2 (1-r)} + 2r -1 & \frac{1}{2 r} + \frac{1}{2(1-r)} -1
\end{array} 
\right) \, . \label{eq:h0qgs}
\end{align}
The NLO and NNLO matrices at the benchmark point (\ref{eq:benchmark}) are 
\begin{align}
\bm{H}^{(1)} &= \left(
\begin{array}{ccc}
 2.73994 & -1.90813 + i\, 0.304853 & 8.37732 - i\, 2.53415 \\ 
-1.90813 - i\, 0.304853 & 1.32846 &  -5.83592 + i\, 0.880136\\ 
8.37732 + i\, 2.53415  & -5.83592 - i\, 0.880136 & 25.6045
\end{array} 
\right) \, , \nn \\
\bm{H}^{(2)} &= \left(
\begin{array}{ccc}
108.917 & -53.1488 -i\,144.053 & 253.200 +i\, 363.288 \\ 
 -53.1488 +i\,144.053& 52.2319 &  -106.493 + i\, 180.767\\ 
253.200 -i\, 363.288 &-106.493 - i\, 180.767  & 597.058
\end{array} 
\right) \, , 
\end{align}
and the anomalous dimension $\bm{\Gamma}$ in this basis is
\begin{align}
\bm{\Gamma} &= \left[\left(N +C_F\right) \gamma_{\text{cusp}} \left(\alpha_s \right) \left(\ln{\frac{s}{\mu^2}} -i \pi \right) + 2 \gamma^g\left(\alpha_s \right) + 2 \gamma^q\left(\alpha_s \right) \right] \bm{1} \nn \\
&+N \gamma_{\text{cusp}} \left(\alpha_s \right) \left( \ln{r} +i \pi \right) \left(
\begin{array}{ccc}
0 & 0 & 0\\ 
0 & 1 & 0\\
0 & 0 & 1
\end{array} 
\right) +  \gamma_{\text{cusp}} \left(\alpha_s \right) \ln\left({\frac{r}{1-r}}\right)\left(
\begin{array}{ccc}
0 & 1 & 0 \\ 
2 &  -\frac{N}{2} & \frac{N^2-4}{2 N} \\
0 & \frac{N}{2}& -\frac{N}{2} 
\end{array} 
\right) \, .
\end{align}
The anomalous dimension $\gamma^g$ can be expanded in powers of $a = \alpha_s/(4 \pi)$ as $\gamma^g = \sum_i a^i \gamma^g_i$. The coefficients of the expansion up to NNLO can be found for example in Appendix~A in \cite{Becher:2013vva}.

\boldmath
\subsubsection{$q(p_1) + g(p_2) \longrightarrow q(p_3)  + g(p_4)$}
\unboldmath

The color basis that we adopt in order to describe the process in~(\ref{eq:qgt}) is 
\begin{align}
{\mathcal C}_1 \equiv  \langle \{a\}| {\mathcal C}_1 \rangle  = \delta^{a_4 a_2} \delta_{a_3 a_1}  \, ,& \qquad&
{\mathcal C}_2 \equiv  \langle \{a\}| {\mathcal C}_2 \rangle  = i f^{a_4 a_2 c} t^c_{a_3 a_1 } \, , \nonumber \\
&{\mathcal C}_3 \equiv  \langle \{a\}| {\mathcal C}_3 \rangle  = d^{a_4 a_2 c} t^c_{a_3 a_1 } &\, . \label{eq:colggt}
\end{align}
The tree-level hard function is 
\begin{align}
\bm{H}^{(0)} = \left( 
\begin{array}{ccc}
\frac{1}{2 N^2} \left(1- r+\frac{1}{1-r}\right) & \frac{1}{N} \left( \frac{3}{2}-\frac{2}{r} - \frac{1}{2 (1-r)}
 - \frac{r}{2}\right) &  \frac{1}{2 N} \left(1- r+\frac{1}{1-r}\right)\\ 
 \frac{1}{N} \left( \frac{3}{2}-\frac{2}{r} - \frac{1}{2 (1-r)}
 - \frac{r}{2}\right)  &  \frac{5}{2}-\frac{r}{2}+ \frac{4}{r^2}- \frac{4}{r}+ \frac{1}{2 (1-r)}&\frac{3}{2} -\frac{2}{r} - \frac{1}{2 (1-r)}
- \frac{r}{2}\\ 
\frac{1}{2 N} \left(1- r+\frac{1}{1-r}\right) & \frac{3}{2}- \frac{2}{r} - \frac{1}{2 (1-r)}
 - \frac{r}{2} &  \frac{1}{2} \left(1-r+\frac{1}{1-r}\right)
\end{array} 
\right) \, , \label{eq:qgtH0}
\end{align}
while the NLO and NNLO matrices at the benchmark point (\ref{eq:benchmark}) 
are
\begin{align}
\bm{H}^{(1)} &= \left(
\begin{array}{ccc}
-0.960743 &  -17.4992 + i\,23.7762& 2.64035 -i\, 6.99918 \\ 
 -17.4992 - i\,23.7762 & 278.010 &  -89.5442 - i\,24.3766\\ 
2.64035 +i\, 6.99918 &  -89.5442 + i\,24.3766 & 24.4888
\end{array} 
\right) \, , \nn \\
\bm{H}^{(2)} &= \left(
\begin{array}{ccc}
466.834  & -1614.79 + i\,37.4939 & 770.946 +i\, 328.609  \\ 
-1614.79 - i\,37.4939 &9379.39 & -3080.95 - i\,2025.95 \\ 
770.946 -i\, 328.609   &-3080.95 + i\,2025.95  & 1148.26
\end{array} 
\right) \, ,
\end{align}
The anomalous dimension $\bm{\Gamma}$ in this basis is
\begin{align}
\bm{\Gamma} &= \left[\left(N +C_F\right) \gamma_{\text{cusp}} \left(\alpha_s \right) \left(\ln{\frac{s}{\mu^2}} -i \pi \right) + 2 \gamma^g\left(\alpha_s \right) + 2 \gamma^q\left(\alpha_s \right) \right] \bm{1} \nn \\
&+\gamma_{\text{cusp}} \left(\alpha_s \right) \left( \ln{r} +i \pi \right) \left(
\begin{array}{ccc}
C_F+N & -1 & 0\\ 
-2 & N-\frac{1}{2 N} & \frac{4-N^2}{2 N}\\
0 & -\frac{N}{2} & N-\frac{1}{2 N} 
\end{array} 
\right) \nn \\
&+\gamma_{\text{cusp}} \left(\alpha_s \right) \ln\left({\frac{r}{1-r}}\right)\left(
\begin{array}{ccc}
0 & 1 & 0 \\ 
2 &  -\frac{N}{2} & \frac{N^2-4}{2 N} \\
0 & \frac{N}{2}& -\frac{N}{2} 
\end{array} 
\right) \, . \label{eq:qguG}
\end{align}

\boldmath
\subsubsection{$q(p_1) + g(p_2) \to g(p_3)  + q(p_4)$}
\unboldmath

The color basis employed in order to describe the scattering process in~(\ref{eq:qgu}) is 
\begin{align}
{\mathcal C}_1 \equiv  \langle \{a\}| {\mathcal C}_1 \rangle  = \delta^{a_3 a_2} \delta_{a_4 a_1}  \, ,& \qquad&
{\mathcal C}_2 \equiv  \langle \{a\}| {\mathcal C}_2 \rangle  = i f^{a_3 a_2 c} t^c_{a_4 a_1 } \, , \nonumber \\
&{\mathcal C}_3 \equiv  \langle \{a\}| {\mathcal C}_3 \rangle  = d^{a_2 a_3 c} t^c_{a_4 a_1 } &\, . \label{eq:colggtu}
\end{align}
The tree-level hard function is 
\begin{align}
\bm{H}^{(0)} = \left( 
\begin{array}{ccc}
 \frac{1}{2 N^2}\left(\frac{1}{r}+r\right) & \frac{1}{N}\left(1-\frac{1}{2 r} - \frac{2}{1-r}+ \frac{r}{2} \right) &  \frac{1}{2 N}\left(\frac{1}{r}+r\right)\\ 
\frac{1}{N}\left(1-\frac{1}{2 r} - \frac{2}{1-r}+ \frac{r}{2} \right)&2 -\frac{4}{1-r}+ \frac{1}{2 r}+ \frac{4}{(1-r)^2}+ \frac{r}{2}  &  1-\frac{1}{2 r} - \frac{2}{1-r}+ \frac{r}{2} \\ 
 \frac{1}{2 N}\left(\frac{1}{r}+r\right) & 1-\frac{1}{2 r} - \frac{2}{1-r}+ \frac{r}{2}  & 
 \frac{1}{2}\left(\frac{1}{r}+r\right)
\end{array} 
\right) \, .
\end{align}
As expected, the matrix above can be obtained from~(\ref{eq:qgtH0}) by replacing $r$ by $1-r$.  The NLO and NNLO matrices at the benchmark point 
(\ref{eq:benchmark}) are
\begin{align}
\bm{H}^{(1)} &= \left(
\begin{array}{ccc}
-2.78334  & -17.7615 +i\, 29.2185 &  -2.76521 -i\,6.64060\\ 
 -17.7615 -i\, 29.2185 & 900.938 & -124.257 - i\, 3.26543 \\ 
-2.76521 +i\,6.64060 & -124.257 + i\, 3.26543 & 8.45881
\end{array} 
\right) \, , \nn \\
\bm{H}^{(2)} &= \left(
\begin{array}{ccc}
568.421 & -3153.14 - i\,6008.72 & 786.213 + i\,1640.19  \\ 
-3153.14 + i\,6008.72 & 34356.2 & -4658.03 - i\, 1019.52 \\ 
786.213 - i\,1640.19 & -4658.03 + i\, 1019.52 & 742.940
\end{array} 
\right) \, .
\end{align}
and the anomalous dimension $\bm{\Gamma}$ in this basis can be obtained by
replacing $r \to 1-r$ in~(\ref{eq:qguG}).

\boldmath
\subsection{Four-gluon scattering}
\unboldmath

For the four-gluon scattering case (\ref{eq:gggg}) we adopt the color
basis used in \cite{Bern:2002tk}, namely
\begin{align}
{\mathcal C}_1 &\equiv  \langle \{a\}| {\mathcal C}_1 \rangle  = 4\mbox{Tr} \left[ t^{a_1} t^{a_2} t^{a_3} t^{a_4}\right] \, ,\nn \\
{\mathcal C}_2 &\equiv  \langle \{a\}| {\mathcal C}_2 \rangle  = 4\mbox{Tr} \left[ t^{a_1} t^{a_2} t^{a_4} t^{a_3}\right] \, ,\nn \\
{\mathcal C}_3 &\equiv  \langle \{a\}| {\mathcal C}_3 \rangle  = 4\mbox{Tr} \left[ t^{a_1} t^{a_4} t^{a_2} t^{a_3}\right] \, , \nn \\
{\mathcal C}_4 &\equiv  \langle \{a\}| {\mathcal C}_4 \rangle  = 4\mbox{Tr} \left[ t^{a_1} t^{a_3} t^{a_2} t^{a_4}\right] \, ,\nn \\
{\mathcal C}_5 &\equiv  \langle \{a\}| {\mathcal C}_5 \rangle  = 4\mbox{Tr} \left[ t^{a_1} t^{a_3} t^{a_4} t^{a_2}\right] \, , \nn \\
{\mathcal C}_6 &\equiv  \langle \{a\}| {\mathcal C}_6 \rangle  = 4\mbox{Tr} \left[ t^{a_1} t^{a_4} t^{a_3} t^{a_2}\right] \, ,\nn \\
{\mathcal C}_7 &\equiv  \langle \{a\}| {\mathcal C}_7 \rangle  = 4\mbox{Tr} \left[ t^{a_1} t^{a_2}\right]  \mbox{Tr} \left[ t^{a_3} t^{a_4}\right] \,  ,\nn \\
{\mathcal C}_8 &\equiv  \langle \{a\}| {\mathcal C}_8 \rangle  = 4\mbox{Tr} \left[ t^{a_1} t^{a_3}\right]  \mbox{Tr} \left[ t^{a_2} t^{a_4}\right] \, , \nn \\
{\mathcal C}_9 &\equiv  \langle \{a\}| {\mathcal C}_9 \rangle  = 4\mbox{Tr} \left[ t^{a_1} t^{a_4}\right]  \mbox{Tr} \left[ t^{a_2} t^{a_3}\right] \,  . \label{eq:ggggcolorbasis}
\end{align}
The color basis in~(\ref{eq:ggggcolorbasis}) is over-complete. The factor of 4 in the r.h.s of (\ref{eq:ggggcolorbasis}) arises from the fact that the authors of~\cite{Bern:2002tk} define their color basis by employing color matrices normalized as $\mbox{Tr} [T^a T^b] = \delta_{ab}$, while we re-express their basis in terms of color matrices with the standard normalization $\mbox{Tr} [t^a t^b] = \delta_{ab}/2$.

The tree-level hard function for the process in~(\ref{eq:gggg}) is 
\be
\bm{H}^{(0)} = \left( 
\begin{array}{cccccc|c}
a & b & c & c & b & a &  \\ 
b & d & e & e & d &  b&  \\ 
c & e & f & f &e  & c &  \bm{0}_{3 \times 6}  \\ 
c & e & f & f & e & c &  \\ 
b & d &  e& e &d  & b &  \\ 
a & b & c & c & b & a &  \\ 
 \hline
 &  &  & \bm{0}_{6 \times 3} &  &  & \bm{0}_{3 \times 3}
\end{array} 
\right) \, ,
\ee
where the elements $a, \cdots, f$ are 
\begin{align}
a &= \frac{1}{r^2}- \frac{2}{r}- 2 r+ r^2+3\, , \nn \\
b &= \frac{1}{r}+ \frac{1}{1-r}+ r- r^2-2\, , \nn \\
c &= \frac{1}{r}- \frac{1}{r^2}- \frac{1}{1-r}+ r-1\, , \nn \\
d &= \frac{1}{(1-r)^2}- \frac{2}{1-r}+ r^2+2\, , \nn \\
e &= - \frac{1}{r}- \frac{1}{(1-r)^2}+ \frac{1}{1-r}- r\, , \nn \\
f &=  1+ \frac{1}{r^2}+ \frac{1}{(1-r)^2}\, .
\end{align}
The NLO hard function for the four-gluon scattering process of~(\ref{eq:gggg}) depends on nine independent functions and has the following structure:
\begin{align}
\bm{H}^{(1)} = \left( 
\begin{array}{ccccccccc}
a_{1} & b_{1} & c_{1} & c_{1} & b_{1} & a_{1} &g_{1} &g_{1} &g_{1} \\ 
b^*_{1} & d_{1} & e_{1} & e_{1} & d_{1} &  b^*_{1} &  h_{1} &h_{1} &h_{1} \\ 
c^*_{1} & e^*_{1} & f_{1} & f_{1} & e^*_{1}  & c^*_{1} &  i_{1} &i_{1} &i_{1}\\ 
c^*_{1} & e^*_{1} & f_{1} & f_{1} & e^*_{1} & c^*_{1} &   i_{1} &i_{1} &i_{1}\\ 
b^*_{1} & d_{1} &  e_{1} & e_{1}  &d_{1}   & b^*_{1}  &  h_{1} &h_{1} &h_{1}\\ 
a_{1} & b_{1} & c_{1} & c_{1} & b_{1} & a_{1} &  g_{1} &g_{1} &g_{1} \\ 
g^*_{1}  & h^*_{1}  &i^*_{1}   & i^*_{1} & h^*_{1}  & g^*_{1}  &0 & 0& 0 \\
g^*_{1}  & h^*_{1}  &i^*_{1}   & i^*_{1} & h^*_{1}  & g^*_{1}  &0 & 0& 0 \\
g^*_{1}  & h^*_{1}  &i^*_{1}   & i^*_{1} & h^*_{1}  & g^*_{1}  &0 & 0& 0 \\
\end{array} 
\right) \, ,
\end{align}
where the non-zero elements at the benchmark point (\ref{eq:benchmark}) are
\begin{align}
a_{1} &= 68.8613 \, , \quad &b_1 &=111.212 + i\,18.1565 \, , \quad &c_1 &= -158.807 - i\,55.4626 \, , \nn \\
d_1 &= 179.607 \, ,  \quad &e_1 &=-256.410 - i\,42.2061\, , \quad &f_1 &=359.541 \, , \nn \\
g_1 &= 24.0246 - i\, 22.3654\, ,\quad &h_1 &=38.8726 - i\,36.1879 \, , \quad &i_1&= -62.8973 +i\,58.5533 \, .
\end{align}
The NNLO matrix has the structure
\be
\bm{H}^{(2)} = \left( 
\begin{array}{ccccccccc}
a_{2} & b_{2} & c_{2} & c_{2} & b_{2} & a_{2} &g_{2} &j_{2} &m_{2} \\ 
b^*_{2} & d_{2} & e_{2} & e_{2} & d_{2} &  b^*_{2} &  h_{2} &k_{2} &n_{2} \\ 
c^*_{2} & e^*_{2} & f_{2} & f_{2} & e^*_{2}  & c^*_{2} &  i_{2} &l_{2} &o_{2}\\ 
c^*_{2} & e^*_{2} & f_{2} & f_{2} & e^*_{2} & c^*_{2} &   i_{2} &l_{2} &o_{2}\\ 
b^*_{2} & d_{2} &  e_{2} & e_{2}  &d_{2}   & b^*_{2}  &  h_{2} &k_{2} &n_{2}\\ 
a_{2} & b_{2} & c_{2} & c_{2} & b_{2} & a_{2} &  g_{2} &j_{2} &m_{2} \\ 
g^*_{2}  & h^*_{2}  &i^*_{2}   & i^*_{2} & h^*_{2}  & g^*_{2}  &p_{2} & p_{2}& p_{2} \\
j^*_{2}  & k^*_{2}  &l^*_{2}   & l^*_{2} & k^*_{2}  & j^*_{2}  &p_{2} & p_{2}& p_{2} \\
m^*_{2}  & n^*_{2}  &o^*_{2}   & o^*_{2} & n^*_{2}  & m^*_{2}  &p_{2} & p_{2}& p_{2} \\
\end{array} 
\right) \, ,
\ee
and the value of the 16 independent elements at the benchmark point
(\ref{eq:benchmark}) is 
\begin{align}
a_2 &= 2106.67 \, , \quad &b_2 &=3196.18 + i\,3422.95 \, , \quad &c_2 &= -4797.70 - i\,4902.02 \, , \nn \\
d_2 &= 5188.15\, ,  \quad &e_2 &=-8129.75 +i\, 692.435 \, , \quad &f_2 &=13732.3 \, , \nn \\
g_2 &= 1930.67 - i\, 5041.68\, ,\quad &h_2 &= 2747.40 - i\,8529.46\, , \quad &i_2&= -3470.79 +i\, 13852.3  \, , \nn \\
j_2 &=-9.86728 +i\, 3871.76 \, ,\quad &k_2 &=-392.448 + i\,5892.80 \, , \quad &l_2&=  1609.60 - i\,9483.43 \, , \nn \\
m_2 &=148.769 -i\,202.044 \, ,\quad &n_2 &= -135.769 -i\, 698.760\, , \quad &o_2&= 1194.28 + i\,1181.94  \, , \nn \\
 & &p_2 &= 1041.49\, . \quad && 
\end{align}
The anomalous dimension $\bm{\Gamma}$ in this basis is 
\begin{align}
\bm{\Gamma} &= \left[2 N\gamma_{\text{cusp}} \left(\alpha_s \right) \left(\ln{\frac{s}{\mu^2}} -i \pi \right) + 4 \gamma^g\left(\alpha_s \right) \right] \bm{1} \nn \\
&+\gamma_{\text{cusp}} \left(\alpha_s \right) \left( \ln{r} +i \pi \right) \bm{M}_1
+\gamma_{\text{cusp}} \left(\alpha_s \right) \ln\left({\frac{r}{1-r}}\right) \bm{M}_2 \, ,
\end{align}
where the matrices $\bm{M}_1$ and $\bm{M}_2$ are
\begin{align}
\bm{M}_1 &=\left(\begin{array}{ccccccccc}
N & 0 & 0 & 0 & 0 & 0 & 0 & 0 & -1  \\ 
0 & N & 0 & 0 & 0 & 0 & 0 & -1 & 0  \\ 
0 & 0 & 2N & 0  & 0 & 0 & 0 & 1 & 1 \\ 
0 & 0 & 0 & 2N & 0  & 0  & 0 & 1 & 1 \\ 
0 & 0 & 0 & 0 & N & 0 & 0 & -1 & 0 \\ 
0 & 0 & 0 & 0 & 0 & N & 0 & 0 & -1 \\ 
-1 & -1 & 0 & 0 & -1 & -1 & 0 & 0 & 0 \\ 
0 & 0 & 1 & 1 & 0 & 0 & 0 & 2N & 0 \\ 
0 & 0 & 1 & 1 & 0 & 0 & 0 & 0 & 2 N
\end{array} 
\right) \, , \nn \\
\bm{M}_2 &=\left(\begin{array}{ccccccccc}
0 & 0 & 0 & 0 & 0 & 0 & 1 & 0 & 1  \\ 
0 & -N & 0 & 0 & 0 & 0 & -1 & 0 & 0  \\ 
0 & 0 & -N & 0  & 0 & 0 & 0 & 0 & -1 \\ 
0 & 0 & 0 & -N & 0  & 0  & 0 & 0 & -1 \\ 
0 & 0 & 0 & 0 & -N & 0 & -1 & 0 & 0 \\ 
0 & 0 & 0 & 0 & 0 & 0 & 1 & 0 & 1 \\ 
1 & 0 & 0 & 0 & 0 & 1 & 0 & 0 & 0 \\ 
0 & -1 & -1 & -1 & -1 & 0 & 0 & -2N & 0 \\ 
1 & 0 & 0 & 0 & 0 & 1 & 0 & 0 & 0
\end{array} 
\right) \, . 
\end{align}
Finally, the tree-level soft function is 
\begin{align}
\bm{\tilde{s}}^{(0)} = \frac{V}{N^2}\left(
\begin{array}{ccccccccc}
C_1 & C_2 & C_2 & C_2 & C_2 & C_3 & N V & -N & N V  \\ 
C_2 & C_1 & C_2 & C_2 & C_3 & C_2 & N V & N V & -N  \\ 
C_2 & C_2 & C_1 & C_3 & C_2 & C_2 & -N & N V & N V \\ 
C_2 & C_2 & C_3 & C_1 & C_2 & C_2 & -N & N V & N V \\ 
C_2 & C_3 & C_2 & C_2 & C_1 & C_2 & N V & N V & -N  \\ 
C_3 & C_2 & C_2 & C_2 & C_2 & C_1 & N V & -N & N V \\ 
N V & N V  & -N  & -N  & N V  & N V  & N^2 V & N^2  & N^2  \\ 
-N & N V  & N V & N V & N V & -N &  N^2 & N^2 V  & N^2 \\ 
N V & -N & N V & N V & -N & N V &  N^2 & N^2 & N^2 V
\end{array} 
\right) \, ,
\end{align}
with $C_1 = N^4 - 3 N^2 +3$, $C_2 = 3 -N^2 $, and $C_3 = 3+N^2$.

\section{Conclusions}
\label{sec:conclusions}
We have given results for the spin-averaged hard functions for all
$2\to 2$ scattering processes in massless QCD up to NNLO in the strong
coupling constant.  These hard functions are a necessary ingredient
for resummations in processes mediated by $2\to2$ scatterings at
Born level, typical examples being dijet and boosted top production.

We extracted our results from NNLO calculations of UV-renormalized
helicity amplitudes presented in \cite{Glover:2003cm,Glover:2004si,
  Bern:2003ck, DeFreitas:2004tk,Bern:2002tk}, using a calculational
procedure explained in Section~\ref{sec:calcproc}.  The main idea is
to interpret the IR poles in the color-decomposed helicity amplitudes
as the UV poles of effective-theory operators, and to subtract them in
the $\msbar$ scheme.  The hard functions defined through this
procedure depend on the basis used in the color decomposition, which
we specified in Section~\ref{sec:results}. In all cases we performed
several non-trivial cross-checks on our (lengthy) results for the
matrix valued, basis-dependent hard functions, which are $2 \times 2$
matrices for four-quark processes, $3 \times 3$ matrices for
two-quark two-gluon processes, and $9 \times 9$ matrices for the
four-gluon process.  We have listed their explicit numerical values at
a benchmark point in Section~\ref{sec:results}, which will facilitate
future cross-checks.  Moreover, we have provided analytic results in
{\tt Mathematica} form with the electronic submission of this paper to
ensure their easy accessibility.  Our results will thus be useful for
practitioners of higher-order resummations in the near and distant
future.

\section*{Acknowledgments}
We are grateful to Nigel Glover for providing us an electronic version
of the results of \cite{Glover:2001af, Glover:2001rd} and to Thomas Becher for several useful discussions concerning the renormalization scheme change presented in \cite{Becher:2013vva}. The work of
A.F. was supported in part by the PSC-CUNY Award No. 66590-00-44 and
by the National Science Foundation Grant  No.~PHY-1417354.
The work of Z.Z. was supported in part by
the National Science Foundation Grant No.~PHY-1068317.

\end{document}